\documentclass[12pt,twoside]{article}
\usepackage{psfig}
\usepackage{amssymb}
\begin{document}
\begin{center}

{\Large\bf QUALITATIVE ANALYSIS\\[5PT]
OF A SCALAR-TENSOR THEORY\\[5PT]
WITH EXPONENTIAL POTENTIAL\\[5PT]}
\medskip
 
{\bf A.B. Batista\footnote{e-mail: brasil@cce.ufes.br},
J. C. Fabris\footnote{e-mail: fabris@cce.ufes.br}, S.V.B. Gon\c{c}alves\footnote{e-mail: svbg@if.uff.br} and
J. Tossa\footnote{e-mail: jtoss@syfed.bj.refer.org. Permanent address:
IMSP, Universit\'e Nationale du B\'enin, R\'epublique du B\'enin.}}  \medskip

Departamento de F\'{\i}sica, Universidade Federal do Esp\'{\i}rito Santo, 
29060-900, Vit\'oria, Esp\'{\i}rito Santo, Brazil \medskip\\
\medskip

\end{center}
\begin{abstract}
\vspace{0.7cm}
A qualitative analysis of a scalar-tensor cosmological model, with an exponential
potential for the scalar field, is performed. The phase diagram for
the flat case is constructed. It is shown that solutions with an initial
and final inflationary behaviour appear. The conditions for which the
scenario favored by supernova type Ia observations becomes an attractor in the space of
the solutions are established.

PACS number(s): 04.20.Cv., 04.20.Me
\end{abstract}
 
\section{Introduction}

The results from the high redshift supernova type Ia observations
\cite{perlmutter,riess} indicate that the Universe is in an
accelerated expansion regime. This means that the matter content
of the Universe must be dominated by an exotic fluid whose pressure
is negative with $p < - \frac{\rho}{3}$. Generally, the cosmological
constant, with $p = - \rho$, is assumed as the most natural candidate
to represent this exotic fluid.
However, the theoretical problems concerning a cosmological constant,
somehow connected with its interpretation as a vacuum energy
\cite{carroll}, make
other possibilities very attractive also.
\par
Some kinds of topological defects can also lead to an effective
equation of state in the searched range. Non-relativistic domain walls,
for example, implies $p = - \frac{2}{3}\rho$. But, a more fashion
theoretical proposal is the so-called quintessence model
\cite{stein}, which is
a minimal coupled scalar field with a slow evolving potential.
It has been argued that such kind of potentials may originate from
supergravity models, and the accelerated solutions correponds to
an attractor \cite{jerome}.
These kind of models may be tracked back to the reference \cite{peebles},
where they have been proposed in order to solve
the dark matter puzzle. Quintessence model makes use of a large
variety of forms for the potential:
exponential\cite{stein}, polynomial combined with exponenential
\cite{jerome}, hyperbolic
sine or cosine \cite{matos} or even a double exponential
\cite{ng}.
\par
In a recent work \cite{sergio}, it has been showed that an exponential potential model
admits as particular solution the typical perfect fluid solution of
the Friedmann-Robertson-Walker model, for any value of the barotropic
equation of state parameter $\alpha$ ($p = \alpha\rho$). Hence, inflationary
power-law solutions are covered by this model. One of our interest here is to
study the status of this particular solutions, verifying to which extent they are attractors,
and if it is possible to obtain models where the expansion is initially
non-accelerated, becoming accelerated later. Due to the fact that
the standard cosmological model needs an inflationary phase in its
primordial phase, it would be also interesting to have models where an
initial and final inflationary phase occur, with an intermediate
non-inflationary behaviour, in order not to spoil the nucleossynthesis
achievements and
structure formation.
\par
In order to do so, we will perform a qualitative analysis of this
model, first in the scalar-tensor model, and secondly with the scalar-tensor
model coupled to ordinary matter. Some studies have already been made
for these cases in the literature \cite{halliwell,coley1,coley2}.
However, in \cite{halliwell} it has been mainly verified that
the flat case is an attractor for a great variety of exponential factor,
while in \cite{coley1,coley2} the analysis, with respect to this
exponential model, was mainly
dedicated to the identification of the nature of critical points.
\par
In the present case, we will be interested in mapping completly the solutions
for $k = 0$, since it has already been shown that $k = 0$ is an attractor with
respect to $k = \pm 1$. Another reason is that
$k = 0$ seems to be favoured by observations \cite{boomerang}. The phase diagram will be constructed and the
positive energy and inflationary regions will be identified.
It comes out that the both kind of desirable models, as
described before, appear and the power-law
particular solutions are indeed attractors of the physical
acceptable solutions.
\par
In next section we describe with some detail the scalar-tensor model with
exponential potential. In section 3, the phase diagram is constructed.
The coupling to ordinary matter is discussed in section 4, while in
section $5$ the conclusions are presented.

\section{The scalar-tensor model}

A minimal coupling between gravity and a self-interacting scalar field
is represented by the lagrangian,
\begin{equation}
\label{mod}
L = \sqrt{-g}\biggr[R - \phi_{;\rho}\phi^{;\rho} + 2V(\phi)\biggl] \quad .
\end{equation}
The field equations are
\begin{eqnarray}
R_{\mu\nu} - \frac{1}{2}g_{\mu\nu}R &=& \phi_{;\mu}\phi_{;\nu} -
\frac{1}{2}g_{\mu\nu}\phi_{;\rho}\phi^{;\rho} + g_{\mu\nu}V(\phi) \quad ,\\
\square\phi &=& - V'(\phi) \quad ,
\end{eqnarray}
where the prime means derivative with respect to $\phi$.
Inserting the FRW metric
\begin{equation}
ds^2 = dt^2 - a^2(t)\biggr[\frac{dr^2}{1 - kr^2} + r^2(d\theta^2 +
\sin^2\theta d\phi^2)\biggl]
\end{equation}
where $k = 0,-1,1$ corresponds to a flat, open and closed Universe
respectively, it results the following equations of motion:
\begin{eqnarray}
\label{em1}
3\biggr(\frac{\dot a}{a}\biggl)^2 + 3\frac{k}{a^2} &=& 
\frac{1}{2}\dot\phi^2 + V(\phi) \quad , \\
\label{em2}
- 2\frac{\ddot a}{a} - \biggr(\frac{\dot a}{a}\biggl)^2 -
\frac{k}{a^2} &=& 
\frac{1}{2}\dot\phi^2 - V(\phi) \quad ,\\
\label{em3}
\ddot\phi + 3\frac{\dot a}{a}\dot\phi &=& - V'(\phi) \quad .
\end{eqnarray}
These equations are connected by the Bianchi identities.
If the potential term is written as
\begin{equation}
\label{pot}
V(\phi) = \frac{2}{3}\frac{1 - \alpha}{(1 + \alpha)^2}\exp{(\mp\sqrt{3(1 + \alpha)}\phi)} \quad ,
\end{equation}
where $- 1 \leq \alpha \leq 1$,
the equations of motion (\ref{em1},\ref{em2}) admit, for the flat case, the solutions
\begin{equation}
\label{ps}
a(t) = a_0t^\frac{2}{3(1 + \alpha)} \quad , \quad \phi(t) =
\pm \frac{2}{\sqrt{3(1 + \alpha)}}\ln t \quad .
\end{equation}
\par
Hence, the potential (\ref{pot}) leads to the usual perfect fluid solutions
of Einstein equation as a particular case. When $\alpha > - \frac{1}{3}$,
these particular solutions represent an expanding, decelarating Universe;
if $\alpha < - \frac{1}{3}$, the scale factor describes an expanding,
accelerating (inflationary) Universe. The limiting case $\alpha = - \frac{1}{3}$ corresponds to $a \propto t$, that is, $\dot a > 0$
and $\ddot a = 0$.
So, the perfect fluid solutions may be mimitized by a scalar-tensor
model with a suitable potential. In \cite{sergio} it was shown that
this scalar-tensor model allows to get rid of instabilities that appears in
the perfect fluid models, at perturbative level
and in the small wavelength
limit, when $\alpha < - \frac{1}{3}$.
\par
However, it is important to know if these particular solutions represent
an attractor of the space of all possible solutions of this model.
Moreover, the solutions (\ref{ps}) describe an Universe that is always
inflationary ($\alpha < - 1/3$) or always non-inflationary ($\alpha > -1/3$).
In a realistic Universe, the behaviour of the scale factor
must change with time. If we accept the inflationary paradigm for
the early Universe and if the results for the value of the decelerating
parameter $q = - \frac{\ddot aa}{\dot a^2}$ obtained through the
measurement of the supernova type Ia are confirmed, then the most
realistic model should have an initial and final inflationary behaviour;
between these two inflationary stages, a non-inflarionary behaviour
must take place, allowing the formation of light elements, through the
primordial nucleosynthesis, and formation of local structure through
the gravitational instability mechanism.
\par
The model described by (\ref{mod}), with the potential given by (\ref{pot}), is
rich enough to
 allow solutions more general than (\ref{ps}). In order to
exploit
 all richness of this model, verifying if realistic models as
described
 before are possible, we will perform a qualitative analysis of
(\ref{em1},\ref{em2}). Compactifying the phase space of all
possible solutions
on the Poincar\'e's
sphere, and delimiting the regions corresponding to the positivity of
the potential $V(\phi)$, as well as
the inflationary type solutions, we will indentify the main features of
the cosmological models resulting from (\ref{em1},\ref{em2},\ref{em3}).

\section{The phase space diagram}

In order to perform a qualitative analysis of the system described
above, we define the new variables
\begin{equation}
x = \dot\phi \quad , \quad y = \frac{\dot a}{a} \quad ,
\end{equation}
and we set $B = \pm \sqrt{3(1 + \alpha)}$ and $V(\phi) = V_0e^{B\phi}$.
Hence, equations (\ref{em1},\ref{em2},\ref{em3}) lead to
the planar, homogenous, autonomous two dimensional system
\begin{eqnarray}
\label{ds1}
\dot x &=& \frac{B}{2}x^2 - 3xy - 3By^2 \quad ,\\
\label{ds2}
\dot y &=& - \frac{1}{2}x^2 \quad ,
\end{eqnarray}
subjected to the condition 
\begin{equation}
V(\phi) = y^2 - \frac{1}{6}x^2 > 0 \quad .
\end{equation}
This system admits in the finite region of the plane($x,y$) an unique
degenerate critical point $x = 0$, $y = 0$. It corresponds to the
Minkowski space. Its eigenvalues are all zero. 
\par
Now, the invariant rays in this plane are characterized by the solutions
$x = \lambda y$. Inserting this relation in (\ref{ds1},\ref{ds2})
it results a third order polynomial relation for $\lambda$:
\begin{equation}
\lambda^3 + B\lambda^2 - 6\lambda - 6B = 0 \Longleftrightarrow
(\lambda + B)(\lambda^2 - 6) = 0 \Longleftrightarrow \lambda_1 = - B\quad,
\quad \lambda_\pm = \pm \sqrt{6} \quad .
\end{equation}
Hence, there are three invariant rays for $B\neq \pm \sqrt{6}$. Two of them are independent of the
value of $B$ while the third one depends on $B$ and, consequently, on
$\alpha$. This third inviariant ray correspond to the
solution (\ref{ps}). Let [XY], [CC'] and [AA'] denote respectively
the rays $x = - By$, $x = - \sqrt{6}y$ and $x = \sqrt{6}y$.
\par
Inserting the expressions for these invariant rays into
(\ref{ds2}) the direction of evolution along them can be determined.
In order to complete the analysis, the critical point at infinity must
be found and their nature studied \cite{conti}. To do this, we project the plan
($x,y$) into the Poincar\'e's sphere, introducing new variables
$z, u, v$, such that $\displaystyle x = \frac{u}{z}, \quad  y =
\frac{v}{z}$ subjected to the condition
 $u^2 + v^2 + z^2 = 1$. The new system
reads:
 \begin{eqnarray}
\label{nds1}
\frac{du}{d\tau} &=& bz - cv \quad , \\
\label{nds2}
\frac{dv}{d\tau} &=& cu - az \quad , \\
\label{nds3}
\frac{dz}{d\tau} &=& av - bu \quad ,
\end{eqnarray}
where $ a = \frac{1}{2}u^2z$, $b = z(\frac{B}{2}u^2 - 3uv - 3Bv^2)$
and $c = - \frac{1}{2}u^3 - \frac{B}{2}u^2v + 3uv^2 + 3Bv^3$.
The new parameter $\tau$ is defined by
the equation (\ref{nds3}). The points at infinity are obtained for
$z = 0$ which corresponds to the equator of the Poincar\'e's
sphere. The critical points are determined and
their nature (repulsive, attractive or saddle points) is 
characterized in the usual way \cite{conti}. The coordinates of these
points are
\begin{eqnarray}
A(u,v,z) = \left(\sqrt{\frac{6}{7}}, \frac{\sqrt{7}}{7}, 0 \right) \quad ,
\quad 
 A'(u,v,z) = \left(- \sqrt{\frac{6}{7}}, - \frac{\sqrt{7}}{7}, 0 \right)
\quad ,\nonumber\\ \nonumber
C(u,v,z) = \left( \sqrt{\frac{6}{7}}, - \frac{\sqrt{7}}{7},0 \right) \quad ,
\quad C'(u,v,z) = \left(- \sqrt{\frac{6}{7}}, \frac{\sqrt{7}}{7}, 0\right)
\quad , \\ \nonumber  X(u,v,z) = \left(
\frac{-B}{\sqrt{1+B^2}},\frac{1}{\sqrt{1+B^2}} ,0 \right) \quad , \quad  
Y(u,v,z) = \left( \frac{B}{\sqrt{1+B^2}},\frac{-1}{\sqrt{1+B^2}} ,0 \right) 
\quad .
\end{eqnarray}
\par
The final phase diagram contain six regions separated by the
three invariant rays. Three of the six regions may be obtained from
the other three by inverting the time: $t \rightarrow - t$. In order to
interprete the behaviour and the nature of the curves,
we complement this
diagram with some other physical considerations.
\par
The complete diagram depends on the value of $B$. However, before
establish these diagrams two physical requirements will be introduced.
The first one concerns the positivity of the contribution of the energy 
of the potential term. The potential $V(\phi)$ is negative in the
region interior to the rays such that $\lambda = \pm \sqrt{6}$. In this region the variable $x$ can never be zero, since
in this case the variable $y$ becomes imaginary.
\par
We will be also interested in identifying the regions where inflation
can occurs. This implies to require $\ddot a > 0$. Hence, $\dot y + y^2
> 0$. Due to (\ref{ds2}), this implies $- \frac{1}{2}x^2 + y^2 > 0$.
Hence, the inflationary region is bounded by the rays $y = \pm \frac{\sqrt{2}}{2}x$.
These rays will be represented by [PP`] and [QQ']. The inflationary regime
occurs in
 the region bounded by these lines and containing the $y$ axis.
\par
Hence, in these phase diagram the two invariant rays (from three), the
condition of positivity of the energy, and the condition to have inflation
are fixed and independent of $B$. Only one invariant ray depends on $B$;
it corresponds to the particular solution (\ref{ps}). When $\alpha = 1$,
this invariant ray coincides with one of other two.
\par
The complete diagram
is displayed in figure 1 for $- \sqrt{2} < B < 0$. As $B$ changes from
$- \sqrt{6}$ to $\sqrt{6}$, the invariant ray [XY] moves from [CC'] to [AA'].
When $\alpha = - 1$, the invariant ray [XY] coincides with the $y$ axis.
\par
If
$- \frac{1}{3} < \alpha < 1$, ($- \sqrt{6} < B < - \sqrt{2}$ or
$\sqrt{2} < B < \sqrt{6}$), the particular perfect fluid-type solutions are non-inflationary.
There are two physically interesting kind of trajectories in this
case. Those connecting the critical point at infinity $C$ to the origin,
approaching asymptotically the invariant ray $[YX]$, start with a free scalar
field
 behaviour ($a \propto t^{1/3}$), and coincide later with the
corresponding
 perfect fluid solution; these curves describe non-inflationary
Universe.
 However, the curves connecting the critical point at infinity $Y$
to the
 origin, start with a non-inflationary perfect fluid behaviour, 
becomes latter inflationary and assymptotically coincides again with
the corresponding non-inflationary perfect fluid solution.
\par
When $ - 1 < \alpha < - \frac{1}{3}$, which corresponds to $- \sqrt{2} < B < \sqrt{2}$, 
the perfect fluid-type particular solutions corresponding to the invariant
ray $[XY]$ are inflationary. Again, we have two kind of interesting
trajectories. The solutions starting from the critical point at
infinity $C$ begins again with a free scalar field, non-inflationary, behaviour and then evolves towards an inflationary regime
as they approach the origin. Those solutions starting from the critical
point $Y$, have initially an inflationary behaviour, becomes later
non-inflationary and again become inflationary as they approach the
origin along the invariant ray $[XY]$. There are also solutions
that begin at $Y$, ending in the origin, remaining always
inflationary.
\par
These last trajectories are the most interesting one since the
standard cosmological model requires an initial inflationary regime,
as well as a final inflationary regime, if the supernova type Ia measurements
are confirmed.
\par
When $B = 0$ ($\alpha = - 1$) all $y$ axis becomes singular and corresponds
to the de Sitter particular solution with different values for the
Hubble factor $\displaystyle \frac{\dot a}{a}$. All physical acceptable
solutions
 go to or come from one point of this singular axis. It is important
to notice that, for any value of $B$, the invariant ray $[XY]$ is
an attractor.

\begin{center}
\begin{figure}
\psfig{file=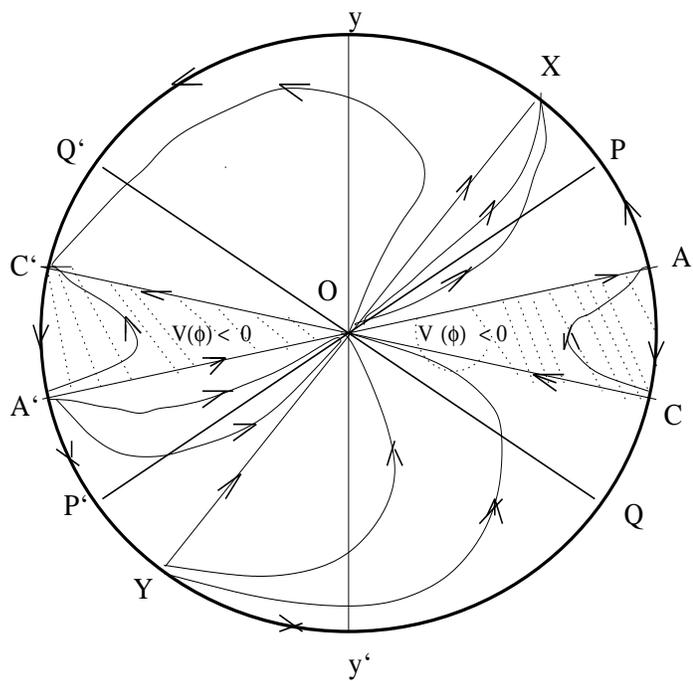,width=18cm,height=15cm,angle=-90.0}
\caption[Phase Diagram of the Dynamical System 2.7]
{\footnotesize \it This diagram gives the Evolution of different solutions
throught inflationary or non inflationary regime. It shows the reheating
phenomena for some solutions in the sector OCX and YOA}
\label{Figure 1}
\end{figure}
\end{center}

\section{Coupling the scalar field to ordinary matter}

The inclusion of ordinary perfect fluid matter, with a barotropic equation,
can be made in a quite direct way. In this case, the equations of motion
read
\begin{eqnarray}
\biggr(\frac{\dot a}{a}\biggl)^2 &=& 8\pi G\rho + \frac{1}{2}\dot\phi^2 + V(\phi) \quad , \\
- 2 \frac{\ddot a}{a} - \biggr(\frac{\dot a}{a}\biggl)^2 &=& 
8\pi G\gamma\rho + \frac{1}{2}\dot\phi^2 - V(\phi) \quad , \\
\ddot\phi + 3\frac{\dot a}{a}\dot\phi &=& - V_\phi(\phi) \quad ,\\
\dot\rho + 3(1 + \gamma)\frac{\dot a}{a}\rho &=& 0 \quad .
\end{eqnarray}
In these expressions, the barotropic equation of state $p = \gamma\rho$
was explicitly employed.
\par
The equations of motion above may be recast in the form of a three-dimensional
dynamic system:
\begin{eqnarray}
\label{ds4}
\dot x &=& Bz + \frac{B}{2}x^2 - 3xy - 3By^2 \quad , \\
\label{ds5}
\dot y &=& - \frac{1 + \gamma}{2}z + x^2 \quad , \\
\label{ds6}
\dot z &=& - 3(1 + \gamma)yz \quad .
\end{eqnarray}
The definition of $x$ and $y$ are the same as before and
$z = 8\pi G\rho$.
For $z = 0$, we go back to the two-dimensional system (\ref{ds1},\ref{ds2}).
\par
The study of a three-dimensional dynamical system is more involved.
However, we may obtain its general features following closely the
procedure employed in the two-dimensional case.
First of all, we remark that the system (\ref{ds4},\ref{ds5},\ref{ds6})
has one critical point, as before, represented by $x = 0$, $y = 0$ and
$z = 0$. It corresponds to the Minkowski space.
But, in order to obtain the invariant rays and the critical points
at infinity, we must perform some suitable "cuts" in the three-dimensional
phase diagram, reducing it to an ensemble of two-dimensional
system which can then be completely analyzed. These "cuts" correspond
to projections of the three-dimensional trajectories on some planes.
\par
The most simple and natural projections correspond to
impose $x = 0$, $y = 0$ or $z = 0$, respectively. The last one
leads to the two-dimensional system analyzed in the previous section.
On the other hand, when we impose $x = 0$ (no scalar field) or $y = 0$
(no gravity), the resulting system can be completly solved.
Specifically, these hypothesis lead to
\begin{eqnarray}
x = 0 \rightarrow y = \sqrt{\frac{z}{3}} \rightarrow a \propto
t^\frac{2}{3(1 + \alpha)} \quad ,\\
y = 0 \rightarrow x = z = 0 \quad (\alpha \geq - 1) \quad .
\end{eqnarray}
The first case corresponds to the gravity perfect fluid system,
whose solution is well known. The second one corresponds to the
trivial case: the Minkowski solution can exist only if matter
and the scalar field are absent also. 
\par
The fact that those two cases are completly solved (one of them
through the trivial solution) just mean that if the solution
is initially in a plane $x = 0$ or $y = 0$, it remains there.
Hence, the most interesting solutions, with non trivial
solutions for $x$, $y$ and $z$, are those outside these planes:
however, their projection on the $z = 0$ behaves as
described in the previous
section, and all analysis performed before remains.

\section{Conclusions}

A self-interacting scalar field coupled to gravity is considered
as a good
candidate to describe the dominant matter content of the Universe today,
leading to an accelerate expansion. The potential term in general is taken
as an exponential function of the scalar field, or a combination of
polynomial and exponential functions. Here we have exploited a
model where the potential is just an exponential function of
the scalar field. In \cite{sergio} it was shown that such potential
term
may lead to power-law solutions for the scale factor typical of
a perfect fluid gravity system, with an arbitrary barotropic equation
of state. Hence, inflationary power-law solutions are included in this
scalar-tensor model.
\par
In the present work, we performed a dynamical analysis of that model
for a flat Universe. It was verified that the power-law particular
solutions act as attractor in the space of the allowble solutions.
Moreover, it was identified,
for some exponential factors,
solutions with an initial and final inflationary
behaviour as well as solutions with just a final inflationary behaviour.
They coincide asymptotically with the particular power-law solutions.
\par
The restriction to $k = 0$ was made due to two reasons mainly:
in \cite{halliwell}, it was shown that in an exponential potential
scalar-tensor model, the flat case is in general an attractor.
Moreover, the CMB anysotropy observations favor a flat Universe
\cite{boomerang}.
The goal of the present work was to present a complete phase diagram
description for this particular, but very important case.
\par
The results showed the richness of the model, and that such a simple
self-interacting scalar field may lead to scenarios that are consistent
theoretically and can be in good agreement with observations.
In particular, if we admit a potential of the
type $V(\phi) = V_0e^{\pm B\phi}$, with $B = 1$, the scale factor
behaves asymptotically as $a \propto t^{2}$, which is one of the
most likely behviours for the scale factor as can be infered from
the supernova type Ia observational programs \cite{efsta}.
\vspace{0.5cm}
\newline
{\bf Acknowledgements:} We thank CNPq (Brazil) for partial
financial support. J.T. thanks the financial support of
CAPES(Brazil) and the Departamento de F\'{\i}sica, Universidade
Federal do Esp\'{\i}rito Santo, for hospitality.

\end{document}